\documentclass[12pt]{article}
\usepackage[dvips]{graphicx}
\def\thefootnote{\fnsymbol{footnote}}
\def\be{\begin{equation}}
\def\ee{\end{equation}}
\def\ba{\begin{eqnarray}}
\def\ea{\end{eqnarray}}

\begin{document}
\begin{titlepage}
\thispagestyle{empty}

\vspace*{2cm}

\begin{center}
{\Large {\bf Order and Chaos}}
\end{center}
\begin{center}
{\Large {\bf in Hofstadter's $Q(n)$ Sequence}}
\end{center}
\vskip0.8cm
\begin{center}
{\large Klaus Pinn}\\
\vskip5mm
{Institut f\"ur Theoretische Physik I } \\
{Universit\"at M\"unster }\\ {Wilhelm--Klemm--Str.~9 }\\
{D--48149 M\"unster, Germany \\[5mm]
 e--mail: pinn@uni--muenster.de
 }
\end{center}
\vskip1.5cm
\begin{abstract}
\par\noindent
A number of observations are made on Hofstadter's integer sequence
defined by $Q(n)= Q(n-Q(n-1))+Q(n-Q(n-2))$, for $n > 2$, and
$Q(1)=Q(2)=1$. On short scales the sequence looks chaotic.  It turns
out, however, that the $Q(n)$ can be grouped into a sequence of
generations. The $k$-th generation has $2^k$ members which have
``parents'' mostly in generation $k-1$, and a few from generation
$k-2$.  In this sense the sequence becomes Fibonacci type on a
logarithmic scale.  The variance of $S(n)=Q(n)-n/2$, averaged over
generations, is $\simeq 2^{\, \alpha \, k}$, with exponent $\alpha =
0.88(1)$.  The probability distribution $p^*(x)$ of $x = R(n)=
S(n)/n^{\alpha}$, $n > \! > 1$, is well defined and strongly
non-Gaussian, with tails well described by the error function erfc. 
The probability distribution of $x_m = R(n)-R(n-m)$ is
given by $p_m(x_m)= \lambda_m \, p^*(x_m/\lambda_m)$, with $\lambda_m
\rightarrow \sqrt{2}$ for large $m$.
\end{abstract}
\end{titlepage}

\setcounter{footnote}{0}
\def\thefootnote{\arabic{footnote}}

\section{Introduction}

In his famous book {\sc G\"odel, Escher, Bach: an Eternal Golden
Braid} \cite{GEB}, Douglas R. Hofstadter introduces a fascinating
integer sequence. In Chapter V he writes:

\begin{quote}
One last example of recursion in number theory leads to a small mystery. 
Consider
the following recursive definition of a function: 
$$
\begin{array}{ll}
 & Q(n) = Q(n-Q(n-1)) + Q(n-Q(n-2)) \quad \mbox{for} ~~ n > 2 \\[5mm]
 & Q(1) = Q(2) = 1 \, .  
\end{array}
$$
It is reminiscent of the Fibonacci definition in that each new value
is a sum of two previous values -- but not of the {\em immediately}
previous two values.  Instead, the two immediately previous values
tell {\em how far to count back} to obtain the numbers to be added to
make the new value! The first 17 $Q$-numbers run as 
follows:\footnote{The outlay of the following 
formula was changed a little by the present
author to avoid typesetting problems.}
$$
1,1,2,3,3,4,5,\underline{5},\underline{6},6,6,8,8,8,10,{\bf 9,10}, \dots 
$$
To obtain the next one, move leftwards (from the three dots)
respectively 10 and 9 terms; you will hit a 5 and a 6, 
indicated by underlining. 
Their sum -- 11 -- yields the new value: $Q(18)$. This is the
strange process by which the list of known $Q$-numbers is used to
extend itself. The resulting sequence is, to put it mildly, erratic.
The further out you go, the less sense it seems to make.  This is one
of those very peculiar cases where what seems to be a somewhat natural
definition leads to extremely puzzling behavior: chaos produced in a
very orderly manner. One is naturally led to wonder whether the
apparent chaos conceals some subtle regularity. Of course, by
definition, there is regularity, but what is of interest is whether
there is another way of characterizing this sequence -- and with luck,
a nonrecursive way.
\end{quote}

\begin{figure}
\begin{center}
\includegraphics[width=11cm]{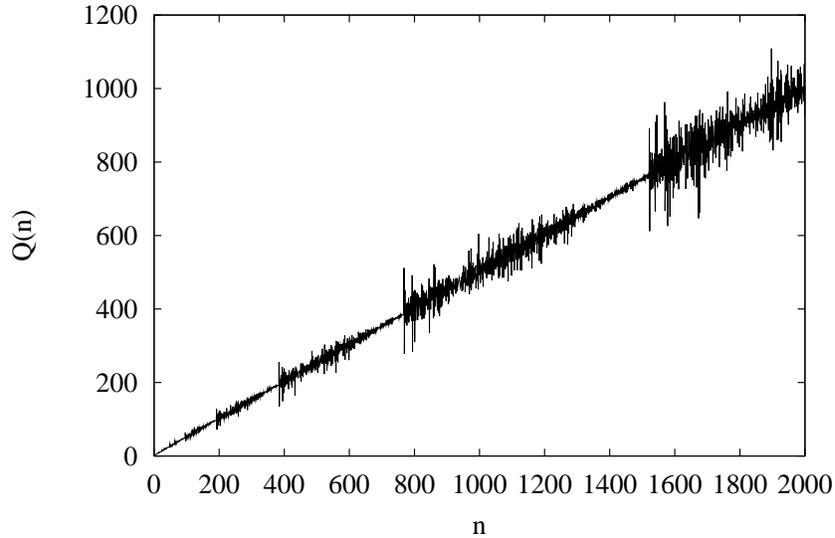}
\parbox[t]{.85\textwidth}
 {
 \caption[appear]
 {\label{appear}
\small
 The first 2000 $Q$-numbers.
 }
 }
\end{center}
\end{figure}

Figure~\ref{appear} gives a first impression of the behavior of the
$Q$-sequence.  It shows the first 2000 members. They scatter around
$n/2$ in a sequence of bursts of increasing amplitude and length.  For
reasons that will become clear later let us call these bursts 
{\em generations}.

Little is known rigorously about the properties of the $Q$-sequence,
though it has found some attention in the literature (see the
discussion by R. K. Guy \cite{Guy}). It has not even been shown that
the sequence is well-defined.

A. K. Yao has done extensive numerical studies \cite{Yao}, mainly
investigating the question of what numbers never appear as values of
the $Q$-function, and in particular if an infinite number of numbers are
left out.  His statistical evidence led him to strongly believe that
an infinite number of values {\em are} left out.

\begin{figure}
\begin{center}
\includegraphics[width=11cm]{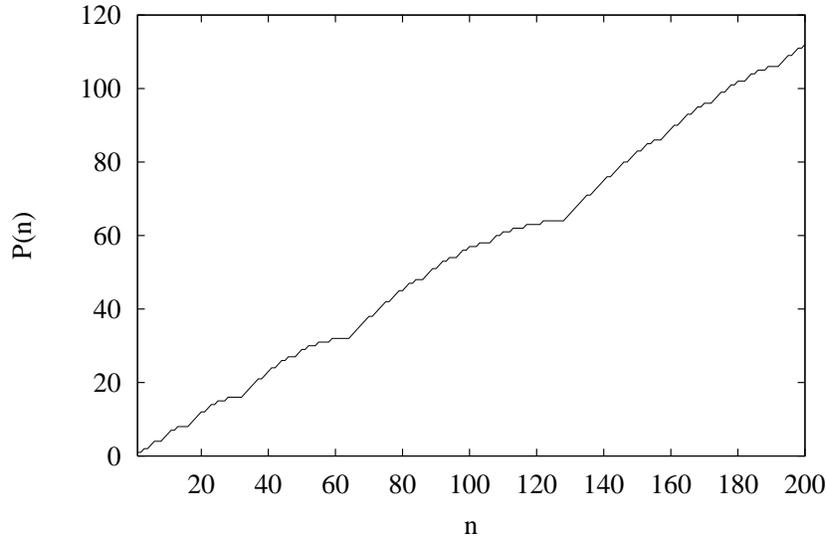}
\parbox[t]{.85\textwidth}
 {
 \caption[appearP]
 {\label{appearP}
\small
 The first 200 $P$-numbers.
 }
 }
\end{center}
\end{figure}

The $Q$-sequence problem inspired some work on related problems, 
e.g.\ on Random Fibonacci-type Sequences \cite{RFS}. 
A well-behaved meta-Fibonacci sequence is the Conway 
sequence 
$$
\begin{array}{ll}
 & P(n) = P(P(n-1)) + P(n-P(n-1)) \quad \mbox{for} ~~ n > 2 \\[5mm]
 & P(0) = P(1) = 1 \, , 
\end{array}
$$
the first 200 elements of which are shown in Figure~\ref{appearP}.
Conway proved that $P(n) \rightarrow n/2$. Mallows \cite{Mallows} won
a cash prize for uncovering the underlying structural properties
of this sequence and establishing its asympotics.\footnote{In fact,
D. R. Hofstadter already invented Conway's sequence and found 
its structure some 10-15 years before Conway posed 
his challenge~\cite{Yao}.}  
Conway's sequence was studied in detail by Kubo and Vakil~\cite{kubo}.

\begin{figure}
\begin{center}
\includegraphics[width=11cm]{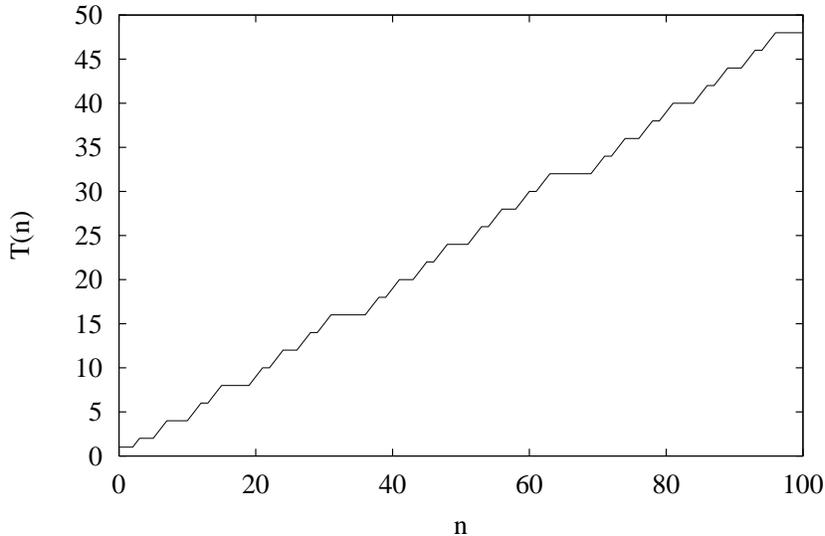}
\parbox[t]{.85\textwidth}
 {
 \caption[appearT]
 {\label{appearT}
\small
 The first 101 $T$-numbers.
 }
 }
\end{center}
\end{figure}

S. M. Tanny studied another sequence, defined through 
$$
\begin{array}{ll}
 & T(n) = T(n-1-T(n-1)) + T(n-2-T(n-2)) \quad \mbox{for} ~~ n > 2 \\[5mm]
 & T(0) = T(1) = T(2) = 1 \, .  
\end{array}
$$
He proved that the $T$-sequence behaves in a completely predictable fashion. 
In particular, $T(n)$ is monotonic and hits every positive integer, cf.\ 
Figure~\ref{appearT}.

In this article I will report on a study of some (mainly statistical)
properties of the $Q$-sequence.  Despite its local irregularity and
chaos, the $Q$-sequence reveals some fascinating structure and order
when looked at on a hierarchy of scales.

\section{Small $n$ Behavior: Parents and Children}

Le us call $Q(n)$ the {\em child} (i.e.\ sum)
of its {\em mother} $Q(n-Q(n-1))$ and {\em father} 
$Q(n-Q(n-2))$.
The arguments $n-Q(n-1)$ and   
$n-Q(n-2)$ will be called the {\em spots} of  
the mother and father, respectively. 

Note that the two parents of a child may be identical, 
i.e.\ live on the same spot $m$ and have the same size $Q(m)$. 
Furthermore, gender does not play a role. 
The notion of parents and children is justified by the observation 
that the $n$'s can be grouped in {\em generations} such 
that children belonging to generation $k$ (with 
some exceptions that seem to be of importance) have parents 
belonging to generation $k-1$. 

\begin{figure}
\begin{center}
\includegraphics[width=11cm]{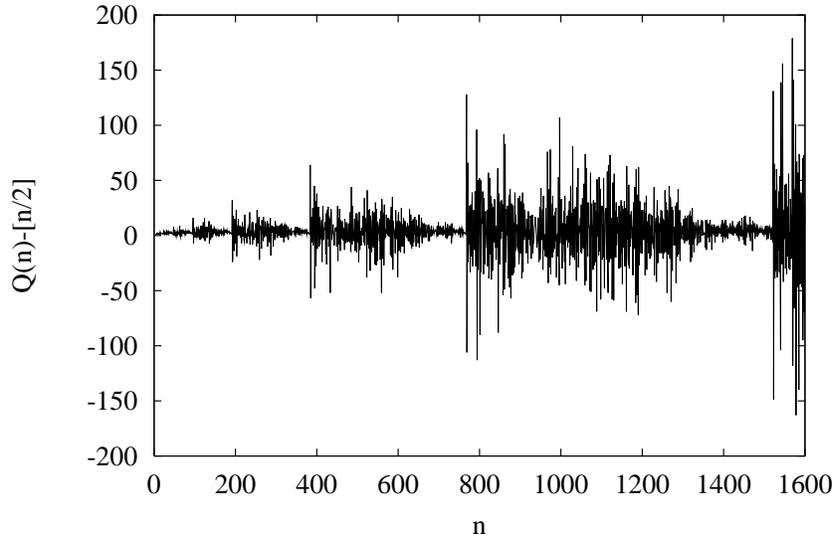}
\parbox[t]{.85\textwidth}
 {
 \caption[burst]
 {\label{burst}
\small
 Regular ``bursts'' in the sequence $S(n)=Q(n)-[n/2]$, 
 located at $n=3,6,12,24,48,96,\dots$ 
 }
 }
\end{center}
\end{figure}

This scenario is suggested already by looking 
at the small $n$ behavior. Figure~\ref{burst} shows the 
sequence $S(n)=Q(n)-[n/2]$, where $[m]$ denotes the integral part
of $m$. ``Bursts'' appear at locations 
$n=3,6,12,24,48,\dots$.
The first large member of a burst is 
always a child of the first
member of the previous burst which is simultaneously
its mother and father. Consequently, it has twice the size 
of its mother-father. The sizes are $Q(3)= 2$, $Q(6)=4$, etc.\ 
Let us call $Q(1)=1$ and $Q(2)=1$ {\em Adam} and 
{\em Eve}. They constitute 
the first generation. The second generation is labelled by
$3,4,5$, the third one starts at $n=6$, and so on. 
An interesting observation is that (most likely similar 
to what happened in human genesis) that Adam has no 
children! Looking carefully at the parenthood relations 
for small $n$, we see that the whole tree is generated
by Eve alone: Her job is to be mother-father of child 3, 
and then together with 3 make 4 and 5 (see Table~\ref{eva}).

\begin{table}
\begin{center}
\begin{tabular}{cccc}
\hline
Generation & $n$  &  &  \\[1mm]
1 & 1    &  (Adam's spot) &  \\  
1 & 2     & (Eve's spot) &   \\[1mm]
\hline
           &      & mother's spot       & father's spot      \\
Generation & $n$  & $n-Q(n-1)$   & $n-Q(n-2)$  \\[1mm]

2 &  3 & 2 & 2 \\
2 &  4 & 2 & 3 \\
2 &  5 & 2 & 3 \\
\hline
3 &  6 & 3 & 3 \\
3 &  7 & 3 & 4 \\
3 &  8 & 3 & 4 \\
3 &  9 & 4 & 4 \\
3 & 10 & 4 & 5 \\
3 & 11 & 5 & 5 \\
\hline
4 & 12 & 6 & 6 \\            
\dots & & & 
 \end{tabular}
\parbox[t]{.85\textwidth}
 {
 \caption[eva]
 {\label{eva}
\small
The first steps in the evolution of $Q(n)$. Adam has no children!
}
}
\end{center}
\end{table}

\begin{figure}
\begin{center}
\includegraphics[width=11cm]{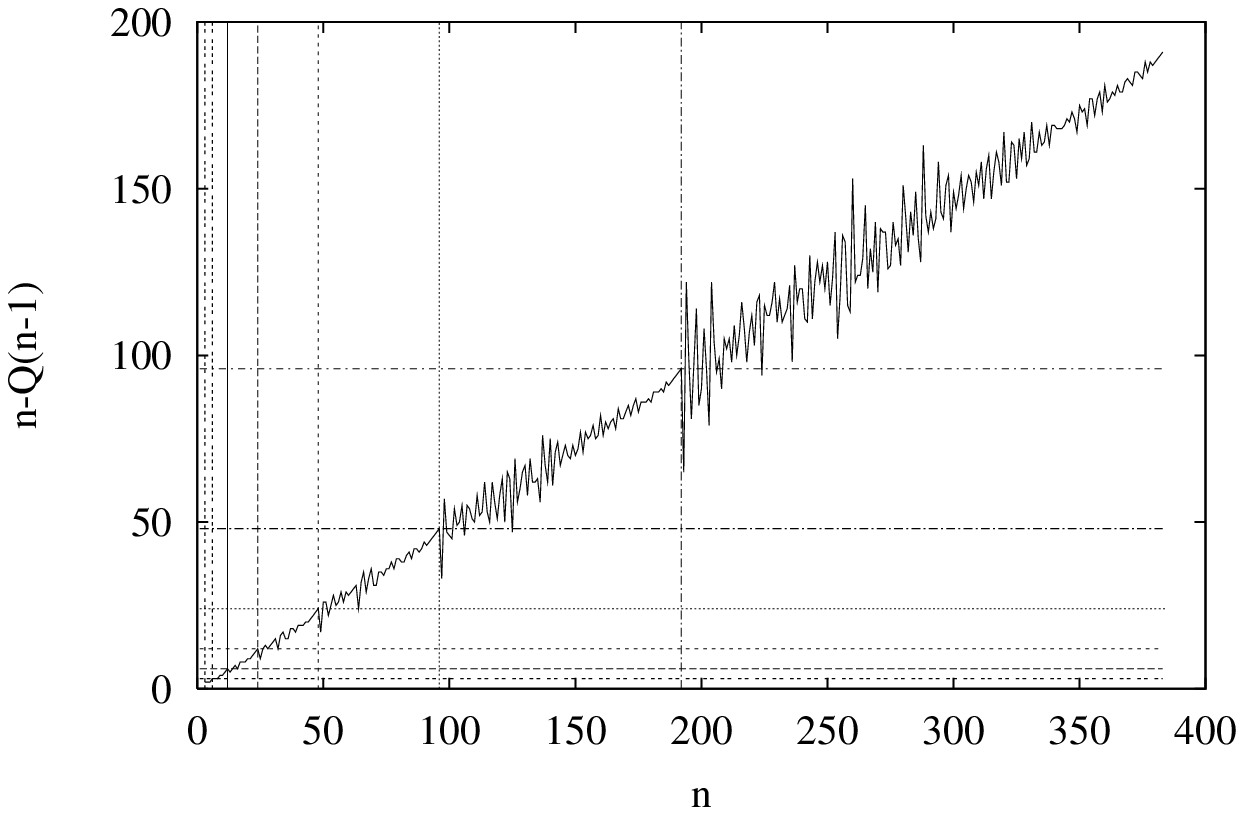}

\includegraphics[width=11cm]{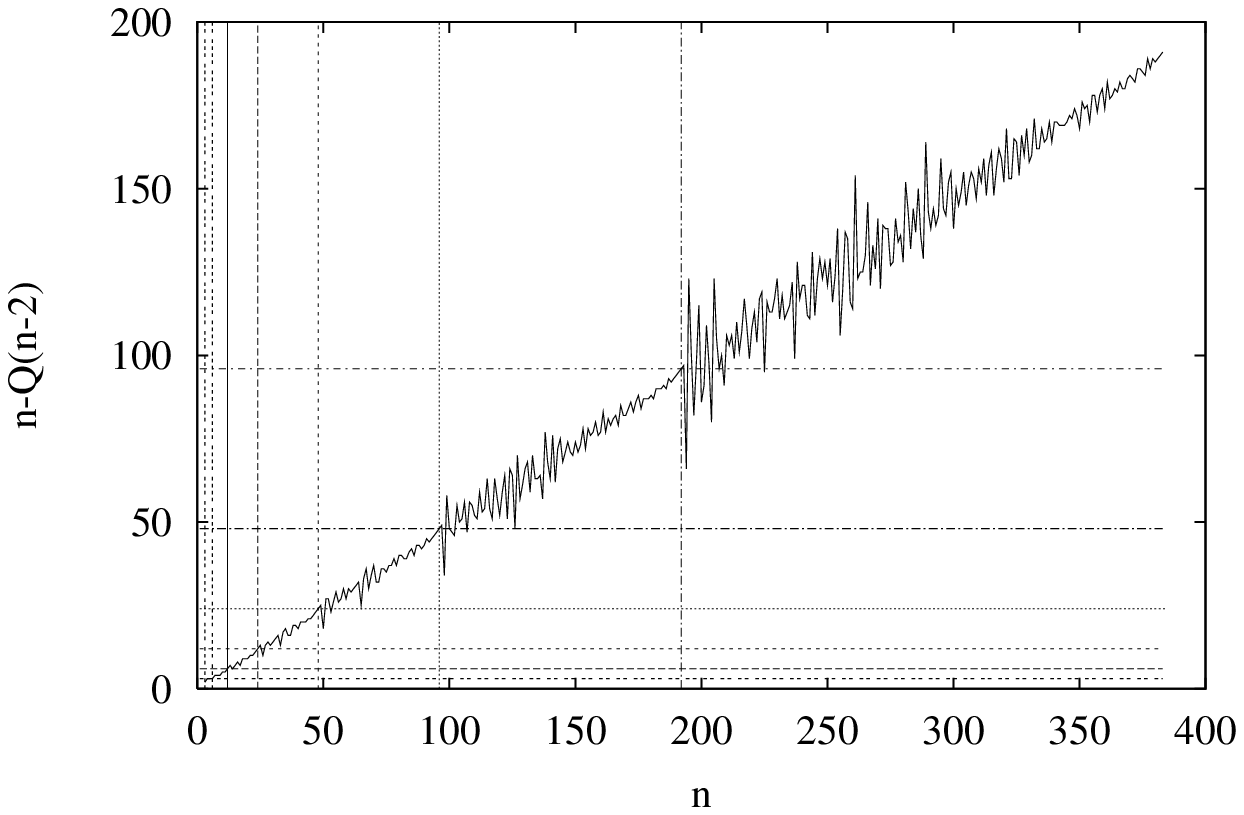}
\parbox[t]{.85\textwidth}
 {
 \caption[pc1]
 {\label{pc1}
Spots $n-Q(n-1)$ and $n-Q(n-2)$
of mothers (top) and fathers (bottom) as function of 
child position $n$, grouped in generations.
 }
 }
\end{center}
\end{figure}

It is important to notice that the parents of the children that
constitute a generation $k$ are mainly in the previous generation.
This is demonstrated in Figure~\ref{pc1}. It shows the spots
$n-Q(n-1)$ and $n-Q(n-2)$ of mothers (top) and fathers (bottom) as
function of child spot $n$, grouped in generations. A careful
inspection reveals that some of the first members of a given
generation get ``genes'' also directly from the next-to-previous
generation. It could be that this fact is relevant for the
observed behavior of the $Q$-sequence.

\section{Behavior for Larger $n$, Exponent $\alpha$}

The strictly regular pattern for the onset of new generations is
broken during the evolution of the $10$th generation starting at
$n=768$. The next burst to follow is located at $n=1522$,
cf.~Figure~\ref{burst}.  Later on the onset of the new generations is
a little less well defined. However, the notion of
generations remains perfectly intact.

\begin{figure}
\begin{center}
\includegraphics[width=12cm]{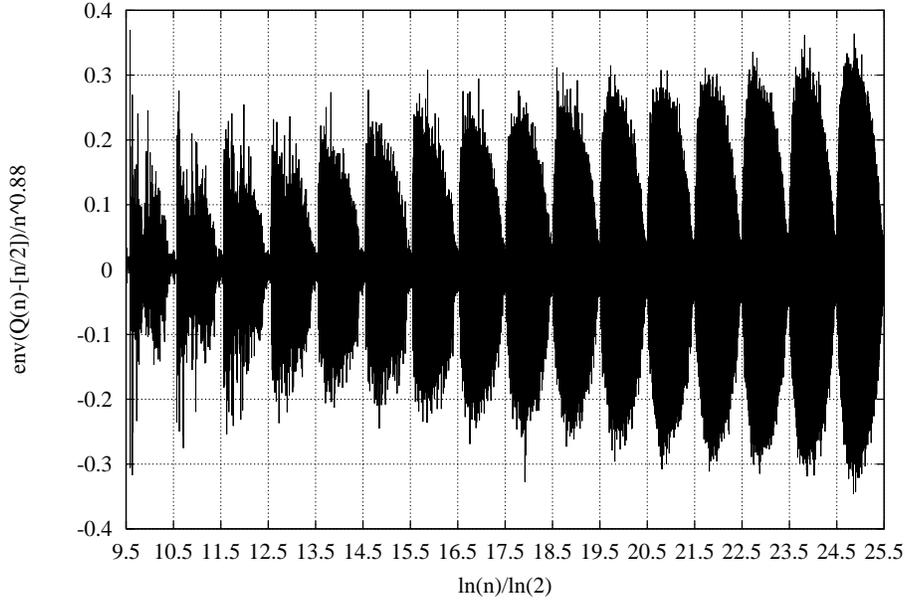}
\parbox[t]{.85\textwidth}
 {
 \caption[env]
 {\label{env}
Envelope of the generations 9 to 24, for $Q(n)-n/2$, 
divided by $n^{\alpha}$, with $\alpha$ = 0.88.
$x$-axis is the logarithm of $n$ with respect to base 2. 
The envelope is obtained by plotting the  minima and maxima 
in intervals of size $\Delta n = n/100$.
}}
\end{center}
\end{figure}

This is demonstrated in Figure~\ref{env}, which shows the generations
9 to 24.  The $x$-axis is the logarithm of $n$ with respect to base 2.
Plotted is the envelope of $Q(n)-n/2$, divided by $n^{\alpha}$, with
$\alpha$ = 0.88.  This power-like rescaling of amplitude will be
discussed in the next section. The envelope is obtained by plotting
the minima and maxima in intervals of size $\Delta n = n/100$.  The
figure clearly shows that the generations populate the intervals
$[2^{k+1/2},2^{k+3/2}]$.

\begin{figure}
\begin{center}
\includegraphics[width=11cm]{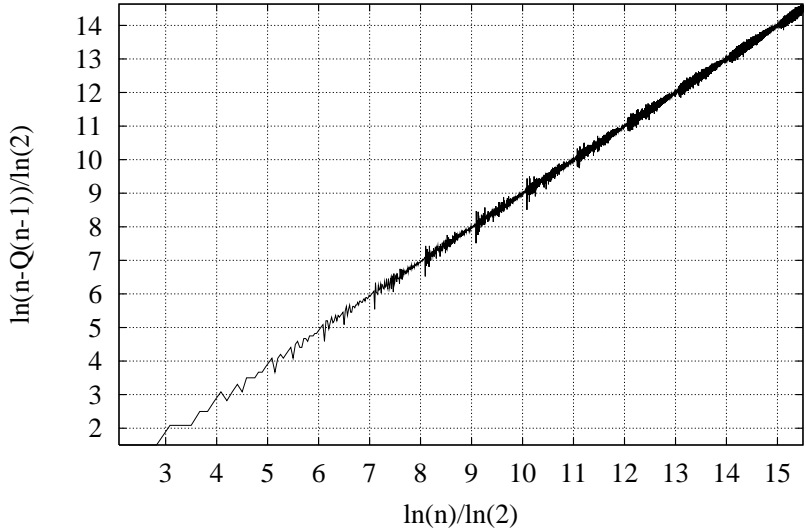}
\parbox[t]{.85\textwidth}
 {
 \caption[pc2]
 {\label{pc2}
Generation of mother vs.\ generation of child. 
 }
 }
\end{center}
\end{figure}

Figure~\ref{pc2} demonstrates that also for large $n$, the mother is nearly
always from the previous generation, sometimes from the
next-to-previous generation, but never older. The same is true 
for the fathers (not plotted).

\section{Rescaling of Amplitude}

We consider the sequence $S(n)= Q(n)-[n/2]$.  Our aim is to compare
the ``size'' of subsequent generations $k$, located in the intervals
$[2^{k+1/2},2^{k+3/2}]$. To this end we define 
a variance $M(k)$ through 
$$
M(k)^2 = 
\bigl\langle 
S(n)^2 
\bigr\rangle_k 
- 
\bigl\langle 
S(n)
\bigr\rangle_k^2 \, , 
$$
where $\langle (.) \rangle_k$ denotes the average over 
the $k$-th generation. 
Table~\ref{msqr} shows numerical results for $ \log_2 M(k)$ 
for generations 8 to 24 and also the quantity
$\log_2 (M(k)/M(k-1))$. 
The results for the latter quantity are fairly constant.
We conclude that 
$$
\frac{M(k)}{M(k-1)} \simeq 2^{\, \alpha} \, , 
$$
with $\alpha = 0.88(1)$.
The variance of the $S(n)$  thus grows in a  
power like fashion, $S(n) \simeq n^\alpha$. 

\begin{table}
\begin{center}
\begin{tabular}{rrc}
$k$ & $\log_2 M(k)$ & $ \log_2 (M(k)/M(k-1))$ \\  
\hline 
    8  &         3.832  &           0.896 \\
   10  &         5.431  &           0.764 \\
   12  &         7.181  &           0.877 \\
   14  &         8.938  &           0.879 \\
   16  &        10.696  &           0.882 \\
   18  &        12.459  &           0.883 \\
   20  &        14.225  &           0.883 \\
   22  &        15.982  &           0.876 \\
   24  &        17.721  &           0.870 \\            
 \hline
 \end{tabular}
\parbox[t]{.85\textwidth}
 {
 \caption[msqr]
 {\label{msqr}
\small
Variances of the generations.
}}
\end{center}
\end{table}

\section{Statistical Distribution Functions}

\begin{figure}
\begin{center}
\includegraphics[width=11cm]{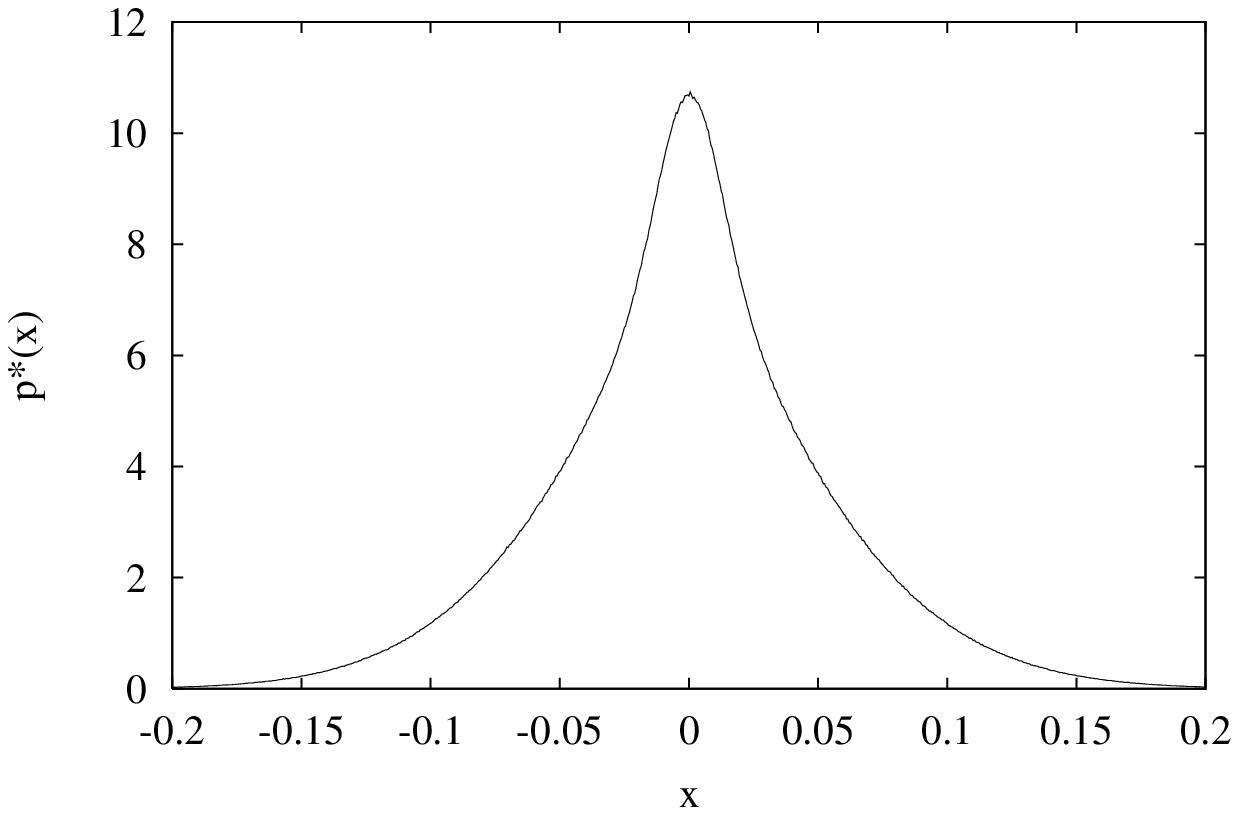}
\includegraphics[width=11cm]{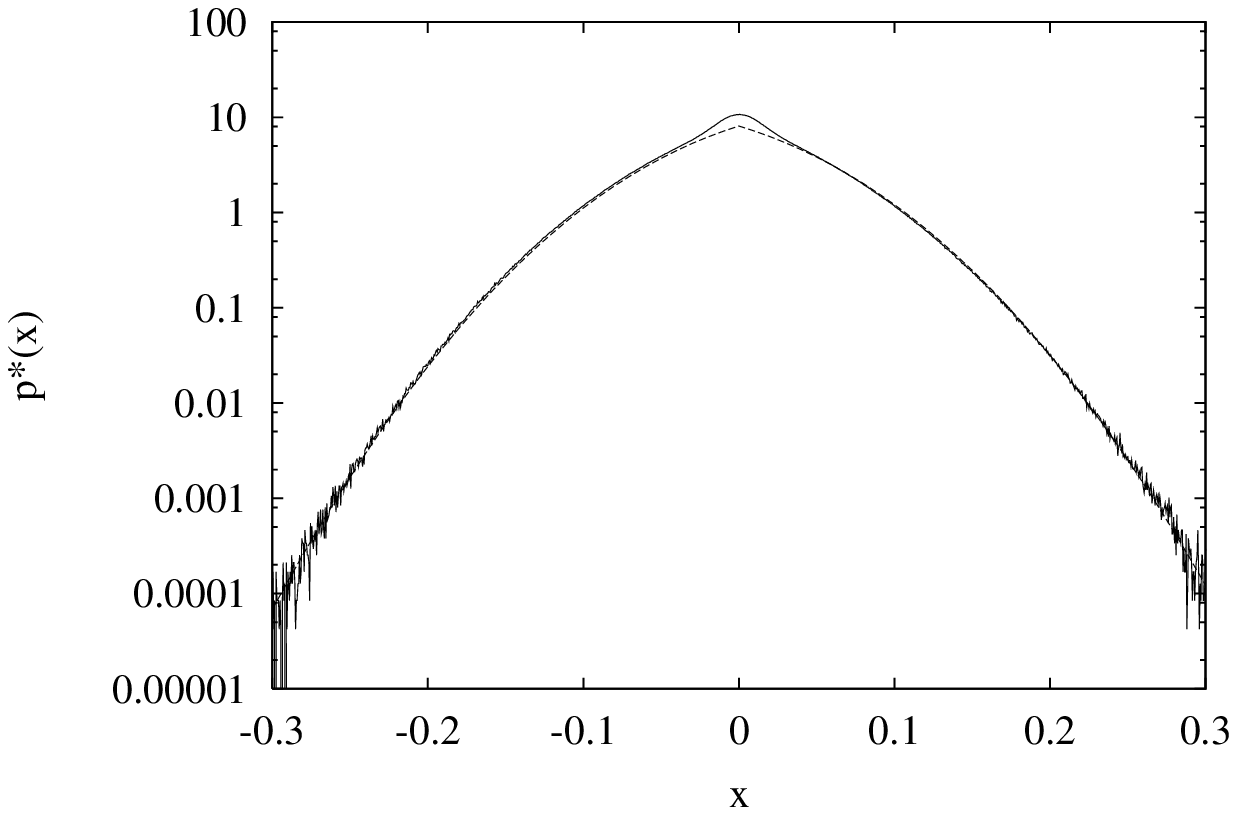}

\parbox[t]{.85\textwidth}
 {
 \caption[bin0]
 {\label{bin0}
Probability distribution of $R(n)$, sampled over the range 
$[2^{13.5},2^{25.5}]$. The lower figure shows $p^*$ 
on a logarithmic scale, together with the functions 
8.1~erfc(-10.5~x) (left wing) and 8.1~erfc(10.2~x) (right wing).
}}
\end{center}
\end{figure}

The previous section suggests that 
$$
R(n)= n^{-\alpha} \, S(n)
$$
could have a well defined probability distribution for large enough
$n$. This is indeed the case.  Figure~\ref{bin0} shows the normalized
histogram of $R(n)$ over the range $[2^{13.5},2^{25.5}]$.  The
distribution, to be called $p^*$, is strongly non-Gaussian. 
The lower part of the 
figure shows $p^*$ on a logarithmic scale, together
with error functions $a \, {\rm erfc}(b  x)$. The parameters 
$a$ and $b$ are specified in the figure caption. The function erfc is 
defined through 
$$
{\rm erfc}(x) = \frac{2}{\sqrt{\pi}}
\int_x^\infty dt \, \exp(-t^2) \, . 
$$
The tails are fitted very well. Note that ${\rm erfc}(x)$ 
decays like $\exp(-x^2)/x$ for large $x$. 

It was confirmed that the distribution was stable against variation of
the sampling range. Furthermore, sampling separately in the
generations yields a sequence of distributions which with increasing
$k$ quickly converge to $p^*$.

\begin{figure}
\begin{center}
\includegraphics[width=12cm]{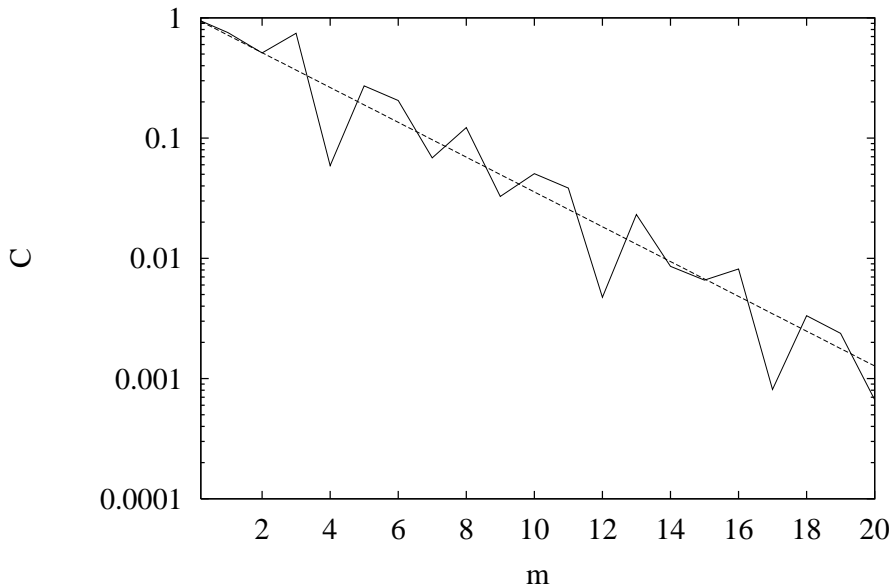}
\parbox[t]{.85\textwidth}
 {
 \caption[lamdec]
 {\label{lamdec}
$ C = |\lambda_m^2 - 2|$, together with the function $\exp(-m/3)$.   
 }}
\end{center}
\end{figure}

A fascinating observation can be made when one looks at the 
distribution $p_m(x_m)$ of differences 
$x_m = R(n)-R(n-m)$. It is 
given by a rescaled $p^*$:
$$
p_m(x_m) = \lambda_m \, p^*(x_m/\lambda_m) \, .
$$
The rescaling factors $\lambda_m$ can be  
computed from the second moments of $x = R(n)$ and 
$x_m = R(n)-R(n-m)$: 
$$
\lambda_m^2 = \frac{ \langle x_m^2 \rangle - \langle x_m \rangle^2 }
                   { \langle x^2 \rangle - \langle x \rangle^2 } \, . 
$$
With increasing $m$ the $\lambda_m$ converge
exponentially to $\sqrt{2}$.
Figure~\ref{lamdec} shows (on logarithmic scale) the quantity 
$ C = |\lambda_m^2 - 2|$, together with the function $\exp(-m/\xi)$, 
with a ``decay length'' $\xi = 3$.

\begin{figure}
\begin{center}
\includegraphics[width=11cm]{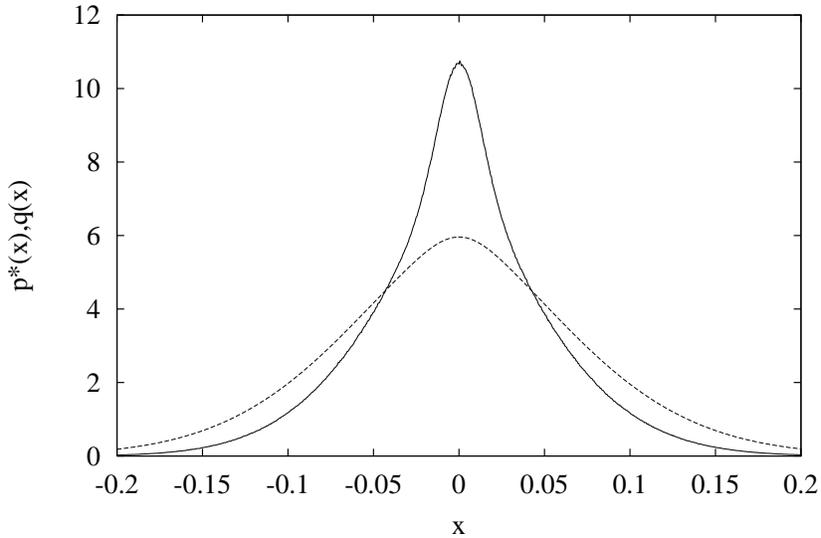}
\parbox[t]{.85\textwidth}
 {
 \caption[conv]
 {\label{conv}
$p^*(x)$ and its self-convolution $q(x)$. 
 }
 }
\end{center}
\end{figure}

Note that this finding implies the existence of long range 
correlations in the $Q(n)$.
Decorrelated $Q$'s would obey a distribution $q$ 
which is given by the convolution of $p^*$ with itself: 
$$
q(x)= \int dy \,  p^*(y) \, p^*(x-y) \, .
$$
Figure~\ref{conv} shows $p^*$ together with its self-convolution. 
The latter already has a close-to-Gaussian shape, and is clearly
different from a rescaled $p^*$.

\section{Conclusions}
The observations reported indicate that the Hofstadter
sequence has a lot of structure and order.
Most likely, many interesting properties of these fascinating
numbers remain to be detected. 
Relations (e.g., by universality) to other systems possessing a 
similar kind of order would be of great interest. 

\section*{Acknowledgements}
I would like to thank D. R. Hofstadter for a  private communication 
and interesting remarks.
Many thanks to P. Grassberger for helpful comments. 
My wife Sabine inspired this study by asking me: ``What does 
it mean to say that the $Q$-sequence is chaotic?''

\end{document}